\documentclass[twocolumn,aps,prb]{revtex4}
\usepackage{graphicx}

\newcommand{\nl}{\nonumber \\}

\newcommand{\be}{\begin{equation}}
\newcommand{\ee}{\end{equation}}
\newcommand{\bea}{\begin{eqnarray}}
\newcommand{\eea}{\end{eqnarray}}

\newcommand{\Eq}[1]{Eq.\,(\ref{#1})}

\newcommand{\la}{\langle}
\newcommand{\ra}{\rangle}
\newcommand{\dg}{\dagger}

\newcommand{\ti}{\tilde}

\begin{document}
\draft

\title{High coherent solid-state qubit from a pair of quantum dots}

\author{Xin-Qi Li$^{1}$ and YiJing Yan$^{2}$}

\address{$^{1}$National Laboratory for Superlattices and Microstructures,
         Institute of Semiconductors,
         Chinese Academy of Sciences, P.~O.~Box 912, Beijing 100083, China \\
         $^{2}$Department of Chemistry, Hong Kong University of Science and Technology,\\
         Kowloon, Hong Kong }

%\date{\today}
\date{February 18, 2002}  %% submitted to Appl. Phys. Lett.

\begin{abstract}
In this letter we propose a scheme to build up high coherent
solid-state quantum bit (qubit) from two coupled quantum dots.
Quantum information is stored in electron-hole pair
state with the electron and hole locating in different dots,
and universal quantum gates involving any pair of qubits are realized
by effective coupling interaction via virtually exchanging cavity photons.
\end{abstract}

\vspace{5ex} 
\maketitle

%%%%%%%%%%%%%%%%%%%%%%%%%%%%%%%%%%%%%%%%%%%%%%%%%%%%%%%%%%%%%%%%%%%%%%%%%%

To be scalable in quantum computation (QC) \cite{Div95255},
recently there have been considerable interest in the
solid-state implementations \cite{Ave98659,Kan98133,Los98120,Ima99}.
%%%%
The main drawback in solid state systems is the severe decoherence.
Thus in the solid-state based QC proposals, the spin (rather than charge) degree of freedom
has been exploited for its relatively long decoherence time \cite{Kan98133,Los98120,Ima99}.
%%%%
On the other hand, some recent impressive experiments investigated
the charge degree of freedom to realize coherent state operation
and entanglement in quantum dots (QDs) \cite{Bon98,Sham00,Haw01},
by making use of the {\it ultrafast} optical spectroscopy
techniques. It is the {\it ultrafast} feature in these experiments
to {\it overcome} the decoherence difficulty.
%%%%
In this letter, we propose a simple scheme to achieve low decoherence qubit from
a pair of weakly coupled QDs.

The basic idea is from our recent work \cite{Li01}, by noticing that
several shortcomings exist there and in some of other QC schemes based on QDs:
(i) In each qubit (two coupled QDs), one and only one excess electron is required
in the conduction band. This is a challenging task within current technology.
(ii) The intersubband transition with THz lasers is currently not a mature technology.
(iii) The coupling between qubits is mediated by Coulomb interactions,
which makes it very difficult to perform conditional gate operation
between any pair of qubits.
%%%%%%%
In this letter, we attempt to remove all these drawbacks based on an all-optical approach.
The physical system we are concerned with is similar
as that proposed by Imamogl\={u} {\it et al} \cite{Ima99,She99}, i.e.,
many qubits (QDs) are located in an optical microcavity, and each qubit can be selectively
performed by laser pulses due to the relatively large distance between qubits.
%%%%%%%%%%%%
However, in our structure, we suggest to use two weakly coupled
QDs (rather than one QD) to construct a qubit as shown in Fig.\ 1,
where only the relevant HOMO and LUMO states are plotted for clarity.
In our present scheme, no excess electron in the
conduction band is required: quantum information is
stored in electron-hole pair state. Namely, logic $|1\ra$
corresponds to electron-hole pair excitation with electron in
state $|e\ra$, and logic $|0\ra$ corresponds to no electron-hole
pair excitation with electron in state $|v\ra$. Since we are
exploiting the charge states for QC, the possible spin
degeneracies (superposition and even decoherence of spin states)
are irrelevant degrees of freedom.

\begin{figure}\label{Fig1}
\begin{center}
\centerline{\includegraphics [scale=0.25] {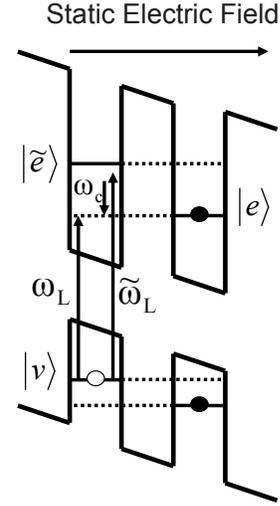}}
\caption{ Schematic diagram for a qubit constructed from two
coupled quantum dots in the presence of external electric field.
The plotted states are resulted from the HOMO and LUMO states of
the individual quantum dots. The ground state $|v\ra$ is used for
logic $|0\ra$, and the excited state $|e\ra$ for logic $|1\ra$.
$|\ti{e}\ra$ plays a role of intermediate state with virtual
occupation. The optical coupling between states are due to the
lasers with frequencies $\omega_L$ and $\ti{\omega}_L$, and cavity
photon with frequency $\omega_C$.  }
\end{center}
\end{figure}

A specific feature we would like to stress is that the energy level structure of the
coupled QDs (nearly identical) under external electric field in
Fig.\ 1 makes the phonon scattering induced decoherence on state
$|e\ra$ be negligible at low temperatures, since the
electron-phonon scattering from $|e\ra$ to $|\ti{e}\ra$ will be
suppressed for a relatively large energy separation (e.g. $\sim
10$ meV) between $|e\ra$ and $|\ti{e}\ra$. In the present QC
scheme, $|\ti{e}\ra$ would play a role of {\it virtually
occupying} intermediate state, thus its decoherence is insensitive to QC.
%%%
Accordingly, the dominant decoherence source in our qubit structure stems from
spontaneous emission of state $|e\ra$, which would be much weaker
than its counterpart in a single dot due to the hole largely locating in another dot.
%%%%%%%%%%%
Regarding the external electric field, its thermal fluctuation (the Johnson noise)
may cause additional dephasing.
To avoid it, the electrodes that generate the electric field
can be connected to a superconductor, which can remove the thermal fluctuations
since there is no dissipation.
%%%%%%%%%%%%%%%%%%%%%%%
In the following, we briefly show how to implement the universal quantum gates,
then present more detailed analysis for decoherence versus operation time.

%%%%%%%%%%%%%%%%%%%%%%%%%%%%%%%%%%%%%%%%%%%%%%%%%%%%%%%%%%%%%%%%%%
%%%%%%%    universal gates    %%%%%%%%%%%%%%%

For single qubit operation, turning on a laser
field with frequency $\omega_L$ on resonance with the energy
difference between $|e\ra$ and $|v\ra$, qubit flipping takes
place under the interaction Hamiltonian
\bea\label{HI1}
H_{I} =  \Omega_L \left[ |e\ra \la v| e^{-i\omega_L t}
       + \mbox{H.c.}  \right] ,
\eea
where $\Omega_L$ is the Rabi frequency. Utilizing this interaction, arbitrary
single-qubit operations can be performed.
%%%%%%%%%%%%%%%%%%%%%%
To realize the universal gates for quantum computation, two-bit
operation such as the controlled-NOT (CNOT), or equivalently,
conditional-phase-shift (CPS) gate, is necessary. To this end, we
suggest to employ the cavity photon to virtually participate in
qubit operation, which can effectively couple the two performed
qubits together.
%%%%%%%%%%%%
Specifically, two lasers selectively act on two qubits (the $j$th and $k$th ones),
both with frequency
$\ti{\omega}_L$ being off-resonance with the transition energy between $|\ti{e}\ra$ and $|v\ra$,
i.e., with detuning $\delta_1=E_{\ti{e}}-E_v-\ti{\omega}_L$.
To avoid real excitation of the cavity photon, $\delta_1$ should slightly differ from
the detuning of the cavity phonon frequency with the level spacing between
$|\ti{e}\ra$ and $|e\ra$, namely, $\delta_1\neq\delta_2=E_{\ti{e}}-E_e-\omega_C$.
%%%%%%%%%%%
Under the laser pulse action,
single qubit (e.g. the $j$th one) effective interaction Hamiltonian reads
\bea\label{HI2}
\ti{H}_{I}
     = \Omega^{(j)}_{\mbox{eff}} \left[ |e\ra_j \la v|a^{\dg}e^{-i\ti{\omega}_Lt}
       + \mbox{H.c.} \right] ,
\eea
with the effective two-photon coupling coefficient
$\Omega^{(j)}_{\mbox{eff}}(t)=\frac{\ti{\Omega}^{(j)}_L(t)\Omega_C}{2}
                         (\frac{1}{\delta_1}+\frac{1}{\delta_2})$,
where $\ti{\Omega}^{(j)}_L$ ($\Omega_C$) is the optical coupling strength between
$|\ti{e}\ra$ and $|v\ra$ ($|e\ra$) due to the laser (cavity photon) field.
%%%%%%%%%%%%%
To implement CNOT or CPS gate, we need to establish a near two-photon
resonance condition between the control ($j$) and target ($k$) qubits,
i.e., $\ti{\delta}^{(j)}=\ti{\delta}^{(k)}\equiv\ti{\delta}=\delta_2-\delta_1$.
Under this condition, and assuming that $\ti{\delta}$ is larger than
$\Omega^{(j)}_{\mbox{eff}}(t)$ and the cavity photon linewidth,
the cavity-photon coordinate in \Eq{HI2} can be eliminated, and
a two-bit effective coupling is established as (in interaction picture with respect to
$H_0=E_e|e\ra\la e|-E_v|v\ra\la v|$ )
\bea\label{HI3}
H_{\mbox{int}}=g_{\mbox{eff}}\left[ \sigma^{+}_{j}\sigma^{-}_{k}
                    + \sigma^{+}_{k}\sigma^{-}_{j}\right],
\eea
where $g_{\mbox{eff}}(t)=\Omega^{(j)}_{\mbox{eff}}(t)
        \Omega^{(k)}_{\mbox{eff}}(t)/\ti{\delta}$.
Here we have introduced the Pauli matrices to denote state transition,
$\sigma^{+}=|e\ra\la v|$ and $\sigma^{-}=|v\ra\la e|$.
Similarly, we can also introduce other Pauli matrices as follows:
$\sigma^{x}=|e\ra\la v|+|v\ra\la e|$,
$\sigma^{y}=-i|e\ra\la v|+i|v\ra\la e|$,
and $\sigma^{z}=|e\ra\la e|-|v\ra\la v|$.
In this way, the present system can be completely mapped to a spin model,
for both the single qubit rotation [c.f. \Eq{HI1}], and two qubit coupling [c.f. \Eq{HI3}].
%%%%%%
With this mapping, the gating technique based on spin interaction
of the form $\sim J {\bf S}^j\cdot{\bf S}^k$ can be
straightforwardly adopted \cite{Los98120,Ima99}. Below we show
that the nontrivial two-bit operation such as the CPS gate can be
realized by combining $H_{\mbox{int}}$ with one-bit operations.
Note that the interaction Hamiltonian of \Eq{HI3} defines a
two-bit {\it joint unitary evolution operator}
$\hat{U}_{jk}(\phi)=T \mbox{exp}[i\int dtH_{\mbox{int}}(t)]$,
where $\phi=\int dt g_{\mbox{eff}}(t)$. With this, the CPS gate can
be implemented using the following pulse sequences \cite{Los98120,Ima99}
\bea\label{CPS}
\hat{U}_{\mbox{CPS}}
   = & & e^{i\pi/4}  e^{i\pi{\bf n}_j\cdot {\bf \sigma}_j/3}
                       e^{i\pi{\bf n}_k\cdot {\bf \sigma}_k/3}
            \hat{U}_{jk}(\pi/4)e^{-i\pi\sigma^y_j/2}                 \nl
     & &\times  \hat{U}_{jk}(\pi/4)
            e^{-i\pi\sigma^x_j/2} e^{-i\pi\sigma^x_k/2}  .
\eea
Here the vector Pauli operator ${\bf \sigma}=(\sigma^x,\sigma^y,\sigma^z)$,
unit vector ${\bf n}_j=(1,1,-1)/\sqrt{3}$, and ${\bf n}_k=(1,-1,1)/\sqrt{3}$.
It can be straightforwardly shown that in the computational subspace
$\{ |vv\ra_{jk},|ve\ra_{jk},|ev\ra_{jk},|ee\ra_{jk} \}$,
$\hat{U}_{\mbox{CPS}}|ee\ra_{jk}=e^{i\pi}|ee\ra_{jk}$, and other basis states
are kept unchanged (with no phase shift).
The more familiar CNOT gate is associated with the above CPS gate in terms of
$\hat{U}_{\mbox{CN}}=\hat{H}^{-1}_k\hat{U}_{\mbox{CPS}}\hat{H}_k$, where
$\hat{H}_k=\mbox{exp}(i\pi\sigma^y_k/4)$ is the Hadarma gate acting on the
$k$th qubit.

%%%%%%%%%%%%%%%%%%%%%%%%%%%%%%%%%%%%%%%%%%%%%%%%%%%%%%%%%%%%%%%%%%%%%%%%%%%%%%%%
%%%%%%%%%%%%%%%%%%%%%%%%%%%%%%%%%%%%%%%%%%%%%%%%%%%%%%%%%%%%%%%%%%%%%%%%%%%%%%%%
%% \section{Operation analysis}

To carry out an analysis for QC operation, we need to specify
the electronic states further.
In two-level approximation, $|e\ra$ and $|\ti{e}\ra$ in Fig.\ 1 are resulted from
coupling of the two isolated dot states
$|d\ra$ and $|\ti{d}\ra$ with coupling strength $t$ and energy separation
$\Delta=E_{\ti{d}}-E_{d}$. (For the highest two valence band states, similar
treatment can be done).
As a result, the eigenstates $|e\ra$ and $|\ti{e}\ra$ have eigenenergies
$E_{\mp}=\frac{1}{2}[(E_d+E_{\ti{d}})\mp\sqrt{\Delta^2+4t^2}]$, and wavefunctions
\bea\label{WFee}
|e\ra &=& \sqrt{1-\gamma}|d\ra+\sqrt{\gamma}|\ti{d}\ra   \nl
|\ti{e}\ra &=& \sqrt{1-\gamma}|\ti{d}\ra-\sqrt{\gamma}|d\ra  ,
\eea
where $\gamma=t^2/(\Delta^2+t^2)$.
With this state nature in mind, we below estimate the decoherence and operation
time in order.

As have mentioned previously, the qubit decoherence time ($\tau_d$) is characterized by
the spontaneous emission rate of $|e\ra$, which is given by Fermi golden rule as
%% \bea\label{W1}
$ 1/\tau_d = \frac{2\pi}{\hbar}\sum_{\bf{q}}|M_{ve}(q)|^2
                          \delta(E_e-E_v-\hbar\omega_q ) $ ,
%%\eea
where $\omega_q$ is the emitted photon frequency,
and $M_{ve}(q)=\la v |H_{\mbox{ep}}(q)|e\ra$ is the perturbative matrix element.
%%%%%%%%%%%%%
For logic gate operation, the time scale ($\tau_G$) is determined
by the optical coupling $H_I^{(L)}$
between $|e\ra$ and $|v\ra$ via the external laser field,
and $H_I^{(C)}$ between $|e\ra$ and $|\ti{e}\ra$ via the cavity photon.
In terms of matrix element of the interaction Hamiltonian,
the coupling strengths can be expressed as
$\Omega_{L(C)}=\la e|H_I^{(L,C)}|v(\ti{e})\ra$.
%%%%%%%%%%%
Due to the spatial separation of state $|e\ra$
from $|v\ra$ and $|\ti{e}\ra$ as shown in \Eq{WFee},
compared to the corresponding counterparts in single dot,
the spontaneous emission rate will be reduced by a factor $\gamma$,
while $\Omega_L$ and $\Omega_C$ will be reduced only by $\sqrt{\gamma}$.
As a consequence, the gate ratio $\rho=\tau_d/\tau_G$ will be enhanced
by a factor $\sim 1/\sqrt{\gamma}$. Similar conclusion has been
quantitatively demonstrated by numerical calculation in Ref.\ \onlinecite{Li01}.

Further, as an order of magnitude estimate, assume $t=0.01$ meV,
and $\Delta=E_{\ti{e}}-E_e=10$ meV.
Accordingly, the {\it spatial separation factor} $\gamma=10^{-6}$.
%%%%%%%%
For the intra-dot interband coupling due to the laser pulse, we
assume $\ti{\Omega}_L=0.1$ meV; and for the
intra-dot state coupling with the cavity photon, the typical value
of $\ti{\Omega}_C=300$ MHz is adopted \cite{She99,Hoo01}.
To avoid real occupation on the state $|\ti{e}\ra$, detuning
$\delta_1=1$ meV is assumed between the laser frequency and the
energy difference between $|\ti{e}\ra$ and $|v\ra$.
%%%%%%%%%
With these parameters, the characteristic time for single qubit rotations
is given by $\tau^{(1)}_{G}=\pi/\Omega_L$, where
$\Omega_L\sim\sqrt{\gamma}\ti{\Omega}_L=10^{-4}$ meV, implying a time
scale of hundreds of nanoseconds.
%%%%
For two qubit operation, the characteristic time $\tau^{(2)}_G$ is dominantly
determined by the two-bit joint evolution [c.f.\ \Eq{CPS}],
$\tau^{(2)}_G\sim \pi/g_{\mbox{eff}}$.
Assuming $\ti{\delta}=\Omega_{\mbox{eff}}/3$, we estimate
$\Omega_{\mbox{eff}}\equiv\Omega^{(j)}_{\mbox{eff}}= \Omega^{(k)}_{\mbox{eff}}
   \simeq \ti{\Omega}_L\Omega_C/\delta_1\simeq 30$ KH,
and $g_{\mbox{eff}}\simeq 10$ KHz.
Accordingly, the two-bit gate such as \Eq{CPS} can be accomplished
within time scale of $10^{-3}$ sec.

For the spontaneous emission, due to the CQED effect and the possible
dark state feature in certain symmetrical QDs \cite{Efr96},
the intra-dot interband radiative lifetime
can be regarded longer than tens to hundreds of microsecond.
%%%%%%%%%%%%%%%%%%%%
Therefore, the qubit decoherence time can be as long as tens of second
(note that $\tau_d\sim\ti{\tau}_d/\gamma \sim 10^6 \times \ti{\tau}_d$),
owing to the spatial separation of the qubit states.
%%%%%%%%%
Within this time scale, the single bit rotation can be performed
as high as $10^8$ times, and the two-bit CNOT gate can
be performed about $10^4$ times.
%%%%%%%%%%%
We notice that the operation quality presented here, say,
the decoherence time and computing speed, is comparable
to the well-known ion-trap QC model \cite{Cir95}.
As an interesting comparison, under the same interaction and relaxation
strengths as assumed above, if the qubit is constructed from single quantum dot,
the coherent operations can be only, $\sim 10^5$ for one-bit rotations,
and $\sim 10$ for the two-bit CNOT gate.

%%%%%%%%%%%%%%%%%%%%%%%%%%%%%%%%%%%%%%%%%%%%%%%%%%%%%%%%%%%%%%%%%%%%%%%%%%
%%%%%%%%%%%%%%%%%%%%%%%%%%%%%%%%%%%%%%%%%%%%%%%%%%%%%%%%%%%%%%%%%%%%%%%%%%
In summary, we have proposed a scheme to build up high coherent solid-state qubit
from a pair of quantum dots, and to implement the univeral quantum gates
by coupling qubits via virtually exchanging cavity photon.
The central challenge to realize the proposed QC scheme is the development
of few-mode THz cavities with extremely low loss. An attractive candidate
is the dielectric cavities, which is currently an intensive research field.

\section*{Acknowledgments}
This work was supported by the special funds for Major State Basic Research Project
No. G001CB3095 of China, and by the special grant from Chinese Academy of Sciences
to distinguished young researchers.

%%%%%%%%%%%%%%%%%%%%%%%%%%%%%%%%%%%%%%%%%%%%%%%%%%%%%%%%%%%%%%%%%%%%%%%%%%

%%%%%%%%%%%%%%%%%%%%%%%%%%%%%%%%%%%%%%%%%%%%%%%%%%%%%%%%%%%%%%%%%%%%%%%%%%


\begin{references}
\bibitem{Div95255}
D.P. DiVincenzo, Science {\bf 269}, 255 (1995);
A. Ekert and R. Josza, Rev. Mod. Phys. {\bf 68}, 733 (1996);
A.M. Steane, Rep. Prog. Phys. {\bf 61}, 117 (1998).
\bibitem{Ave98659}
D.V. Averin, Solid State Commun. {\bf 105}, 659 (1998);
Y. Makhlin, G. Sch\"on, and A. Shnirman, Nature {\bf 398}, 305 (1999);
L.B. Ioffe, V.B. Geshkenbein, M.V. Feigelman, A.L. Fauchere,
and G. Blatter, Nature {\bf 398}, 679 (1999);
J.E. Mooij, T.P. Orlando, L. Levitov, L. Tian, C.H. van der Wal, and S. Lloyd,
Science {\bf 285}, 1036 (1999).
\bibitem{Kan98133}
B.E. Kane, Nature {\bf 393}, 133 (1998).
\bibitem{Los98120}
D. Loss and D.P. DiVincenzo, Phys. Rev. A {\bf 57}, 120 (1998).
\bibitem{Ima99}
A. Imamoglu, D. D. Awschalom, G. Burkard, D. P. DiVincenzo,
D. Loss, M. Sherwin, and A. Small, Phys. Rev. Lett. {\bf 83}, 4204 (1999).
\bibitem{She99}
M. S. Sherwin, A. Imamoglu, and T. Montroy, Phys. Rev. A {\bf 60}, 3508 (1999).
%%%%%  QD exp.  %%%%%
\bibitem{Bon98}
N.H. Bonadeo etal, Phys. Rev. Lett. 81, 2759 (1998); Science 282, 1473 (1998).
\bibitem{Sham00}
G. Chen, N.H. Bonadeo, D.G. Steel, D. Gammon, D.S. Katzer, D. Park, L.J. Sham,
Science {\bf 289}, 1906 (2000).
\bibitem{Haw01}
M.Bayer, P. Hawrylak, K. Hinzer, S. Fafard, M. Korkusinski,
Z.R. Wasilewski, O. Stern, and A. Forchel, Science {\bf 291}, 451 (2001).
%%%%%%%%%%%%%%%%%
\bibitem{Li01}
X.Q. Li and Y. Arakawa, Phys. Rev. A {\bf 63}, 012302 (2001).
\bibitem{Efr96}
Al.L. Efros, M. Rosen, M. Kuno, M. Nirmal, D.J. Norris, and M. Bawendi,
Phys. Rev. B {\bf 54}, 4843 (1996).
\bibitem{Hoo01}
C.J. Hood, H.J. Kimble, and J. Ye, e-print quant-ph/0101103.
\bibitem{Cir95}
J.I. Cirac and P. Zoller, Phys. Rev. Lett. {\bf 74}, 4091 (1995).


\end{references}
\end{document}